\documentstyle[twocolumn,prl,aps,epsf]{revtex}
\begin{document}

\draft
\wideabs{
\title{Tunable Molecular Resonances of Double Quantum Dots Embedded in an
 Aharonov-Bohm Interferometer}
\author{Kicheon Kang$^1$ and Sam Young Cho$^2$}
\address{   $^1$Basic Research Laboratory, Electronics and Telecommunications
            Research Institute, Taejon 305-350, Korea}
\address{   $^2$Department of Physics, University of Queensland, Brisbane 
            4072, Australia }

\date{\today}

\maketitle

\begin{abstract}
We investigate resonant tunneling through molecular states of coupled double
quantum dots embedded in an Aharonov-Bohm (AB) interferometer.
The conductance through the system consists of two
resonances associated with the bonding and the antibonding
quantum states. We predict that the two resonances are 
composed of a Breit-Wigner resonance and a Fano resonance,
those widths and Fano factor depending on the AB phase very sensitively.
Further, we point out that the bonding properties, such as the covalent
and the ionic bonding, can be identified by the AB oscillations.
\end{abstract}
}
%

While single quantum dots are regarded as artificial 
atoms due to their quantization of energies~\cite{kastner93,kouwen01}, 
two (or more) quantum dots can be coupled to form an 
{\em artificial molecule}~\cite{wiel02}.
Resonant tunneling through series quantum dots provides some
informations on the coupling between dots~\cite{vaart95}, but
the phase coherence of the bonding cannot be directly addressed
in this geometry.
Aharonov-Bohm (AB) interferometers containing a 
quantum dot enables investigating the phase coherence
of resonant tunneling through a quantum 
dot~\cite{yacoby95,schuster97,kobayashi02}. 
The phase coherence of the Kondo-assisted transmission has been also studied
in this 
geometry~\cite{wiel00,ji00,gerland00,kang00,bulka01,hofstetter01,kang02}.
Recently, an AB interferometer
setup containing two coupled quantum dots has been 
realized~\cite{holleitner01}. This can be considered as the
beginning point of the study for experimentally unexplored region where
various aspects of double dot molecule can be investigated 
with probing the phase coherence.
There are some previous theoretical works for the AB interferometer containing
two quantum dots.
Resonant tunneling~\cite{kubala02}, cotunneling~\cite{akera93}, 
Kondo effect~\cite{izumida97}, and magnetic polarization current~\cite{cho02}
have been the subjects of the study for the system without direct coupling
between dots.
Two-electron entanglement in the presence of direct tunneling between dots
has been also studied~\cite{loss00} in the context of quantum 
communication.

In this Letter, we study phase-sensitive 
molecular resonances through double quantum dots
embedded in an Aharonov-Bohm interferometer. 
The geometry we consider is schematically
drawn in Fig.~1, and is basically equivalent to the experimental setup of
Ref.~\onlinecite{holleitner01}. 
We find that the conductance through the system consists of two
molecular resonances associated with the bonding and the antibonding
quantum states. By careful analysis on the conductance as a function
of energy, we argue that
the two resonances are {\em always}
composed of a Breit-Wigner resonance and a Fano resonance,  
those widths and Fano factor depending on the AB phase very sensitively. 
Further, we propose that the bonding properties, such as the covalent
and the ionic bonding, can be characterized by the AB oscillations.

Our model is described by the following Hamiltonian:
\begin{mathletters}
\label{eq:hamil}
\begin{equation}
 H = H_M + H_0 + H_T \;,
\end{equation}
where $H_M$, $H_0$, and $H_T$ stand for the artificial molecules of
double quantum dots, two electrical leads, and tunneling between 
the leads and the
quantum dots, respectively. 
For the molecule, we consider coupled non-interacting quantum dots
of energies $\varepsilon_1$, $\varepsilon_2$ with tunneling matrix element
$t$ between two dots,
\begin{equation}
 H_M = \varepsilon_1 d_1^\dagger d_1
  + \varepsilon_2 d_2^\dagger d_2
  - t (d_1^\dagger d_2 + d_2^\dagger d_1) \;,
\end{equation}
where $d_i$ ($d_i^\dagger$) with $i=1,2$ annihilates (creates) an electron
in the $i$-th dot. The bonding properties depend on the
ratio of the energy difference of the quantum dot levels ($\Delta\varepsilon
\equiv \varepsilon_1-\varepsilon_2$) and the
tunnel splitting ($2t$). The molecular bonding can be called `covalent' for
$|\Delta\varepsilon| \ll 2t$ where the eigenstates of the electrons
are delocalized. On the other hand, the molecule is considered to be
in the `ionic' bonding limit for $|\Delta\varepsilon| \gg 2t$ where
the eigenstates are localized in one of the two dots~\cite{wiel02}. 
$H_0$ describes the two (left and right)
electrical leads modeled by the Fermi sea as,
\begin{equation}
 H_0 = \sum_{k\in L} E_k^L a_k^\dagger a_k
     + \sum_{k\in R} E_k^R b_k^\dagger b_k \;,
\end{equation}
where $a_k$ ($a_k^\dagger$) and $b_k$ ($b_k^\dagger$) annihilates (creates)
an electron in the left and one in the right leads, respectively. 
These two leads are assumed to be identical, 
$E_k\equiv E_k^L = E_k^R$.
Finally, tunneling between the leads and the molecule is described by
\begin{equation}
 H_T = -\sum_{k,i=1,2} (V_L^i d_i^\dagger a_k + \mbox{\rm H. c.})
       -\sum_{k,i=1,2} (V_R^i d_i^\dagger b_k + \mbox{\rm H. c.}) \;. 
\end{equation}
\end{mathletters}
For simplicity, we assume that the magnitudes of tunneling matrix 
elements for the four different arms are same, denoted by $V$.
Then the matrix elements 
can be written as 
$V_L^1=V_R^2= Ve^{i\varphi/4}$, $V_L^2=V_R^1= Ve^{-i\varphi/4}$. 
The phase factor $\varphi$ comes from the AB
flux, and is defined by $\varphi = 2\pi\Phi/\Phi_0$ where $\Phi$ and
$\Phi_0$ are the external flux through the interferometer 
and the flux quantum ($=hc/e$), respectively.
The hopping strength between a quantum dot and a lead is denoted
by $\Gamma$, defined as 
\begin{equation}
 \Gamma = 2\pi\rho(E_F) V^2 \;,
\end{equation}
where $\rho(E_F)$ stands for the density of states of each leads
at the Fermi level, $E_F$.

The Hamiltonian is transformed by using symmetric and antisymmetric
modes of the leads and the quantum dots. Later it will become obvious that
this approach provides better insights into the problem. Let us consider
the transformations of electron operators 
\begin{mathletters}
\begin{eqnarray}
 \alpha_k &=& (a_k + b_k)/\sqrt{2}, \;\; \beta_k = (a_k - b_k)/\sqrt{2}, \\
 d_\alpha &=& (d_1 + d_2)/\sqrt{2}, \;\; d_\beta = i(d_1 - d_2)/\sqrt{2}. 
\end{eqnarray}
\end{mathletters}
Note that, for $\varepsilon_1=\varepsilon_2$,
$d_\alpha$ and $d_\beta$ correspond to the annihilation operator
of the bonding and the antibonding modes. 
By adopting this transformation we can rewrite the Hamiltonian as follows:
\begin{mathletters}
\label{eq:hamil_t}
\begin{equation}
 H = H_\alpha + H_\beta + H_{\alpha\beta} \;,
\end{equation}
where $H_\alpha$, $H_\beta$ take the simple form of the Fano-Anderson 
Hamiltonian~\cite{mahan90} ($\gamma=\alpha,\beta$)
\begin{equation}
 H_\gamma = \tilde{\varepsilon}_\gamma d_\gamma^\dagger d_\gamma
   + \sum_k E_k \gamma_k^\dagger \gamma_k
   + V_\gamma \sum_k (d_\gamma^\dagger\gamma_k + \gamma_k^\dagger d_\gamma) 
 \;,
\end{equation} 
where the energy eigenvalues 
of the two `quantum dot' modes
are given by $\tilde{\varepsilon}_\alpha=\varepsilon_0 - t$, 
$\tilde{\varepsilon}_\beta
=\varepsilon_0 + t$ where $\varepsilon_0 = (\varepsilon_1+\varepsilon_2)/2$.  
The hybridization matrix elements depend
on the AB phase as $V_\alpha = -2V\cos{(\varphi/4)}$ and 
$V_\beta = -2V\sin{(\varphi/4)}$.
The coupling between two modes is given by
\begin{equation}
 H_{\alpha\beta} = -\bar{t} d_\alpha^\dagger d_\beta 
    -\bar{t}^* d_\beta^\dagger d_\alpha \;, \label{eq:h_ab}
\end{equation}
with the `tunneling' matrix element being proportional to the difference
of the energy levels of two quantum dots, $\bar{t} = i(\Delta\varepsilon)/2$.
\end{mathletters}
It is important to note that the coupling term given in Eq.(\ref{eq:h_ab})
vanishes for $\varepsilon_1=\varepsilon_2$. In other words, for the same
single particle energies of the two dots, the original Hamiltonian 
is mapped onto the problem of two independent Fano-Anderson Hamiltonian.

In the representation of the transformed Hamiltonian (Eq.(\ref{eq:hamil_t}))
the dimensionless conductance can be written as~\cite{meir92}
\begin{equation}
 {\cal G} = \frac{1}{4} |\Gamma_\alpha G_\alpha(E_F)
     - \Gamma_\beta G_\beta(E_F)|^2 
     + \Gamma_\alpha\Gamma_\beta |G_{\alpha\beta}(E_F)|^2 \;,
\end{equation}
where $\Gamma_\alpha$, $\Gamma_\beta$ stand for the hopping strengths
between the discrete level and the continuum
of each modes given by $\Gamma_\alpha = 2\Gamma\cos^2{(\varphi/4)}$,
$\Gamma_\beta = 2\Gamma\sin^2{(\varphi/4)}$, respectively.
$G_\alpha(E_F)$ ($G_\beta(E_F)$) and $G_{\alpha\beta}(E_F)$ denote the
diagonal and the off-diagonal components of the $2\times2$ Green's function
matrix. After some algebra for the Green's function we can obtain a very
compact form of the conductance
\begin{mathletters}
\label{eq:cond}
\begin{equation}
 {\cal G} = 
   \frac{ (e_\beta-e_\alpha)^2 + 4\Delta }{ | (-e_\alpha+i)(-e_\beta+i)
        - \Delta |^2 } \;,
\end{equation}
where 
\begin{eqnarray}
 e_{\alpha,\beta} &\equiv& \frac{2}{\Gamma_{\alpha,\beta}} 
   (\tilde{\varepsilon}_{\alpha,\beta} - E_F) \;, \\
 \Delta &\equiv& \frac{4|\bar{t}|^2}{\Gamma_\alpha\Gamma_\beta} 
   = \frac{(\Delta\varepsilon)^2}{\Gamma_\alpha\Gamma_\beta} \;. 
\end{eqnarray}
\end{mathletters}
Note that Eq.(\ref{eq:cond}) becomes equivalent to the one obtained in
\onlinecite{kubala02} in the absence of direct coupling between
two quantum dots ($t=0$).

First we discuss the covalent bonding limit, $\varepsilon_1=\varepsilon_2$. 
This limit
is very instructive to give insights into the problem, since the coupling
term in the Hamiltonian~(\ref{eq:hamil_t}) vanishes. Therefore,
$H_{\alpha\beta}=0$,
and it is clear that the transport is associated with two 
resonances of widths $\Gamma_\alpha$ and $\Gamma_\beta$.  
In this limit the conductance reduces to
\begin{equation}
 {\cal G} = \frac{ (e_\beta-e_\alpha)^2 }{ (e_\alpha^2+1)(e_\beta^2+1) } \;.
  \label{eq:cond0}
\end{equation}

In the following we argue that the conductance consists of the convolution
of a Breit-Wigner resonance and a Fano resonance of the two molecular 
(the bonding and the antibonding) states for $\varepsilon_1=\varepsilon_2$. 
Since the hopping parameters 
$\Gamma_\alpha$, 
$\Gamma_\beta$ are sensitive to the AB phase, one can manipulate
the resonance widths via AB flux. Let us consider the limit 
$\Gamma_\alpha \gg \Gamma_\beta$. For the energy scale larger 
than $\Gamma_\beta$ ($|e_\beta| \gg 1$), 
the conductance of Eq.(\ref{eq:cond0}) follows the Breit-Wigner form of
its width $\Gamma_\alpha$:
\begin{equation}
 {\cal G} \simeq {\cal G}_{BW} = \frac{1}{e_\alpha^2+1} \;. 
  \label{eq:BW}
\end{equation}
On the other hand, one can find that the conductance shows the Fano-resonance
behavior near the antibonding state ($|e_\beta| \lesssim 1$),
\begin{equation}  
 {\cal G} \simeq {\cal G}_{Fano} 
    = {\cal G}_b \frac{(e_\beta+q)^2}{e_\beta^2+1} \;,
   \label{eq:fano}
\end{equation}
where the Fano factor $q$ and the background conductance ${\cal G}_b$ are
given by $q=4t/\Gamma_\alpha$ and ${\cal G}_b=1/(q^2+1)$, respectively. 
From this
analysis we find that the conductance is composed of a Breit-Wigner resonance
around the bonding state ($\tilde{\varepsilon}_\alpha$) with its resonance
width $\Gamma_\alpha$ and a Fano resonance around the antibonding
state ($\tilde{\varepsilon}_\beta$) of width $\Gamma_\beta$, 
if $\Gamma_\alpha\gg\Gamma_\beta$.
Further information is obtained from the Fano factor $q$. For $t=0$ 
the conductance shows an antiresonance behavior ($q=0$),
while it becomes more Breit-Wigner-like (large $q$) 
as the inter-dot coupling increases.
For the other limit $\Gamma_\alpha\ll\Gamma_\beta$, the same analysis
is applied with the role of the bonding and the antibonding states 
interchanged and the Fano factor given by $q=-4t/\Gamma_\beta$.
That is, for $\Gamma_\alpha\ll\Gamma_\beta$, 
the conductance consists of the Breit-Wigner 
resonance for the antibonding state and the Fano resonance near the bonding
state.

Fig.~2 shows the conductance as a function of the Fermi energy. 
The curve demonstrates the features described above, 
which follows the Breit-Wigner and the Fano asymptotes 
for the larger and for the smaller width of the resonances, respectively. 
For $\varphi=0.3\pi$ used in Fig.~2, $\Gamma_\alpha\gg
\Gamma_\beta$, therefore, the conductance shows Breit-Wigner behavior
for the bonding state ($E_F\simeq\tilde{\varepsilon}_\alpha$), 
and Fano resonance around the antibonding state
($E_F\simeq\tilde{\varepsilon}_\beta$). One can also verify that the Fano
factor ($q$) increases as the bonding becomes stronger.
The resonance around the antibonding state evolves from the anti-resonance
for weak bonding (small $q$) to the Breit-Wigner like resonance behavior
for strong bonding (large $q$).

Next, we investigate the more general case 
where $\varepsilon_1\ne\varepsilon_2$.
The same kind of analysis about the resonances can be applied here, with the
Fano resonance modified. We discuss the limit $\Gamma_\alpha\gg\Gamma_\beta$,
without loosing generality. For the energy scale larger than $\Gamma_\beta$,
the conductance takes the Breit-Wigner form of Eq.(\ref{eq:BW}) as in the
$\varepsilon_1=\varepsilon_2$ case. 
However, the conductance near the narrower resonance
($|e_\beta|\lesssim1$) is modified as
\begin{mathletters}
   \label{eq:fano2}
\begin{equation}
 {\cal G} \simeq {\cal G}_{Fano}'
    = {\cal G}_b \frac{|e_\beta'+Q|^2}{e_\beta'^2+1} \;,
\end{equation}
where 
\begin{equation}
 e_\beta'=(e_\beta+{\cal G}_b\Delta q)/(1+{\cal G}_b\Delta) \;,
\end{equation}
and the modified Fano factor given by
\begin{equation}
 Q = q\frac{ 1-{\cal G}_b\Delta }{ 1+{\cal G}_b\Delta }
   + i \frac{2\sqrt{\Delta}}{ 1+{\cal G}_b\Delta } \;.
\end{equation}
\end{mathletters}
Eq.(\ref{eq:fano2}) can be regarded as a generalized Fano resonance formula
with the {\em complex} Fano factor $Q$.
As pointed out in Ref.~\onlinecite{clerk01}, the Fano factor is 
complex number in the absence of the time reversal symmetry, for e.g., by
applying external magnetic field. This point
was addressed experimentally with an AB interferometer containing a
single quantum dot~\cite{kobayashi02}.  

Two significant changes are found in Eq.(\ref{eq:fano2}) compared to
Eq.(\ref{eq:fano}). (i) Since the modified Fano factor $Q$ is a complex
number in general, transmission zero does not exist for $\varepsilon_1 \ne
\varepsilon_2$ unlike the covalent bonding limit. (ii) The width of the Fano
resonance becomes broader due to the difference of the energy levels between
the dots as
\begin{equation}
 \Gamma_\beta' = \Gamma_\beta + {\cal G}_b 
  \frac{ (\Delta\varepsilon)^2 }{ \Gamma_\alpha }\;. 
\end{equation} 
For a fixed value of $\Delta\varepsilon$, one can find that
the broadening of the 
resonance is significant for small inter-dot coupling, recalling the relation 
${\cal G}_b = 1/(q^2+1)$ with $q=4t/\Gamma_\alpha$.

The conductance as a function of the Fermi energy 
for the case of different energy levels is shown in Fig.~3. 
Since $\Gamma_\alpha\gg\Gamma_\beta$ for $\varphi=0.3\pi$, the conductance
shows again the Breit-Wigner resonance behavior for the larger energy
scale corresponding to the ``bonding" state. The modified Fano-resonance
can be observed around the ``antibonding" state. The imaginary
part of the modified Fano factor increases as the bonding strength becomes
stronger, which implies that the resonance around the antibonding state
evolves from the antiresonance to the Breit-Wigner like resonance. 
One can also verify that the width of the Fano resonance is broader here
compared to the case of $\varepsilon_1=\varepsilon_2$.

Finally, we discuss on the Aharonov-Bohm oscillations of the conductance
for the molecular states, and show that the bonding properties are 
identified by the oscillation patterns. 
As shown in Fig.~4, the oscillation patterns
are very different for the ionic and for the covalent bonding limit.
Most of all, the periodicity changes from $2\pi$ in the ionic bonding
to $4\pi$ in the covalent bonding limit. This periodicity change can be
interpreted in terms of the effective coupling strength between two dots. 
In the ionic bonding
limit of $|\Delta\varepsilon| \gg 2t$, the AB oscillation
has the usual $2\pi$-periodicity since the coupling between the dots
is ineffective. However, in the covalent bonding limit, the coupling
between dots becomes important, and this strong effective 
coupling separates the
interferometer into two sub-regions with their cross-sectional area halved.
Therefore the oscillation period is doubled.

Comparing the AB oscillations of the bonding (Fig.~4(b))
and the antibonding (Fig.~4(c)) states, one finds that there are phase
difference by $2\pi$ between the corresponding eigenstates.
This originates from the difference of the wave function symmetry of
the two eigenstates. 

In conclusion, 
we have investigated resonant tunneling through the molecular states of 
coupled two quantum dots embedded in an Aharonov-Bohm interferometer.
We have found that the two resonances 
are composed of a Breit-Wigner resonance and a Fano resonance,
those widths and Fano factor depending on the AB phase.
Further, we have suggested that that the bonding properties and their
symmetries can be characterized by the AB oscillation.



%
\begin{figure} 
\begin{center}
\epsfxsize=2.2in
\epsffile{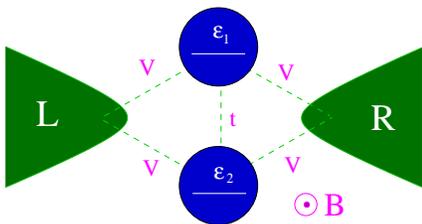}
 \caption{Schematic diagram of double quantum dots embedded in an 
  Aharonov-Bohm interferometer. 
    }
\end{center}
  \label{fig1}
\end{figure}
\begin{figure} 
\epsfxsize=2.8in
\epsffile{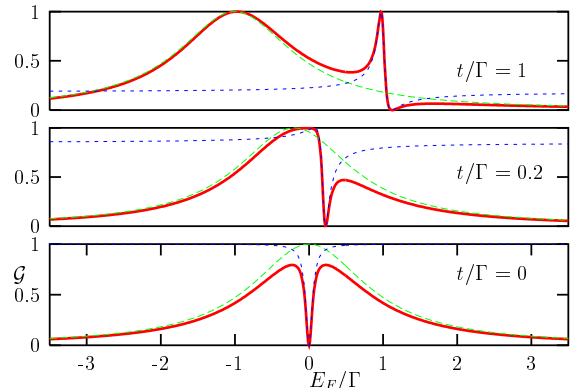}
 \caption{Dimensionless conductance ($\cal G$) as a function of the 
  Fermi energy (solid lines) for three different values of the inter-dot
  tunneling. Other parameters are given by  
  $\varepsilon_1=\varepsilon_2=0$, $\varphi=0.3\pi$.
  Long and short dashed
  lines denote the Breit-Wigner and the Fano asymptotes given in 
  Eq.(\ref{eq:BW}) and Eq.(\ref{eq:fano}), respectively. 
  The Fano factors for the Fano asymptotes are given by
  $q=0$, $q=0.423$, $q=2.115$ for $t/\Gamma=0,0.2,1$, respectively. 
    }
  \label{fig2}
\end{figure}
\begin{figure} 
\epsfxsize=2.8in
\epsffile{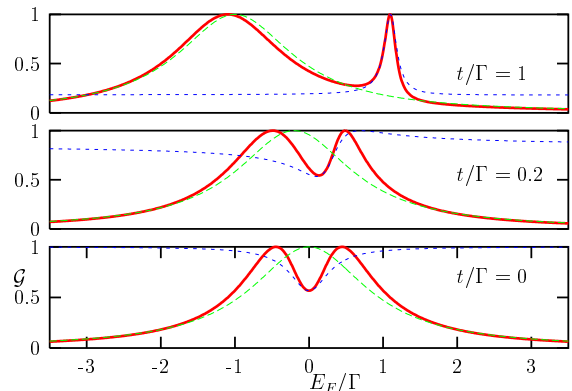}
 \caption{Dimensionless conductance ($\cal G$) as a function of the
  Fermi energy (solid lines) for three different values of the inter-dot
  tunneling. Other parameters are given by $\varepsilon_1=-0.5\Gamma$,
  $\varepsilon_2=0.5\Gamma$, $\varphi=0.3\pi$.
  Long and short dashed
  lines denote the Breit-Wigner and the generalized 
  Fano asymptote given in
  Eq.(\ref{eq:BW}) and Eq.(\ref{eq:fano2}), respectively. The generalized 
  Fano factors for the asymptotes are given by
  $Q=0.753i$, $Q=-0.258+0.861i$, $Q=0.128+2.335i$ for $t/\Gamma=0,0.2,1$, 
  respectively.
    }
  \label{fig3}
\end{figure}
\begin{figure}
\begin{center}
\epsfxsize=2.0in
\epsffile{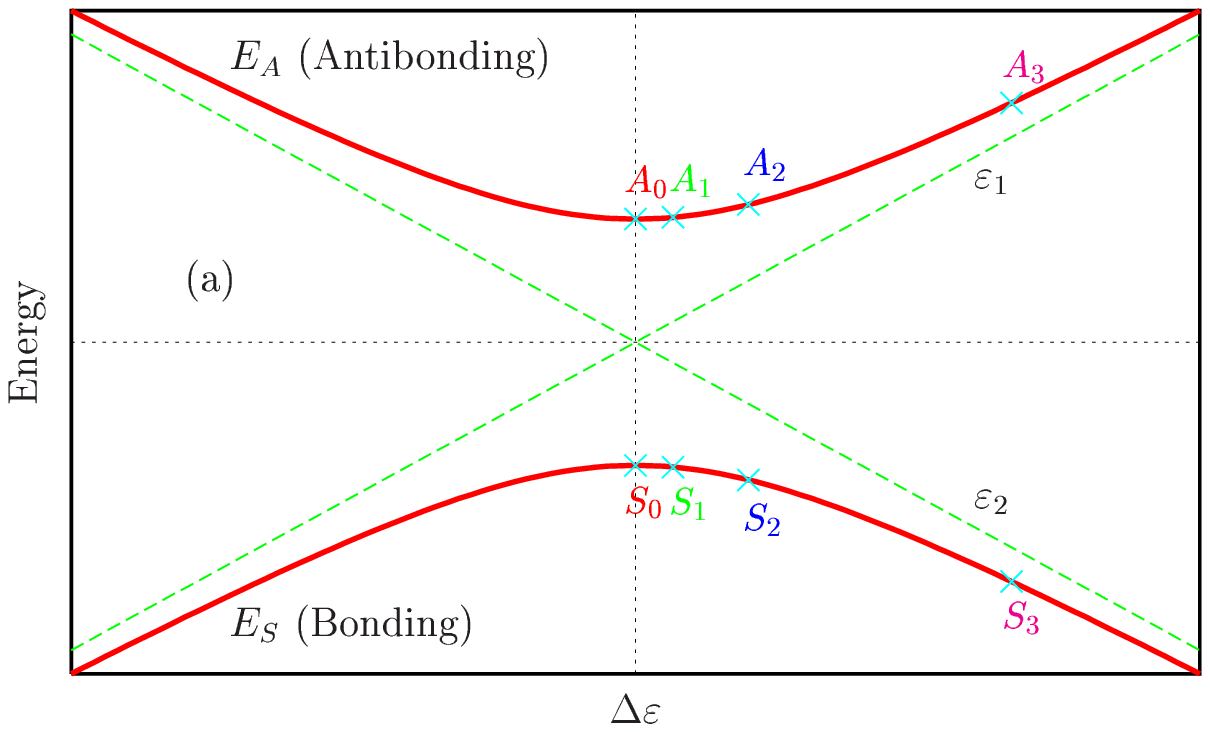}
\end{center}

\epsfxsize=1.6in
\epsffile{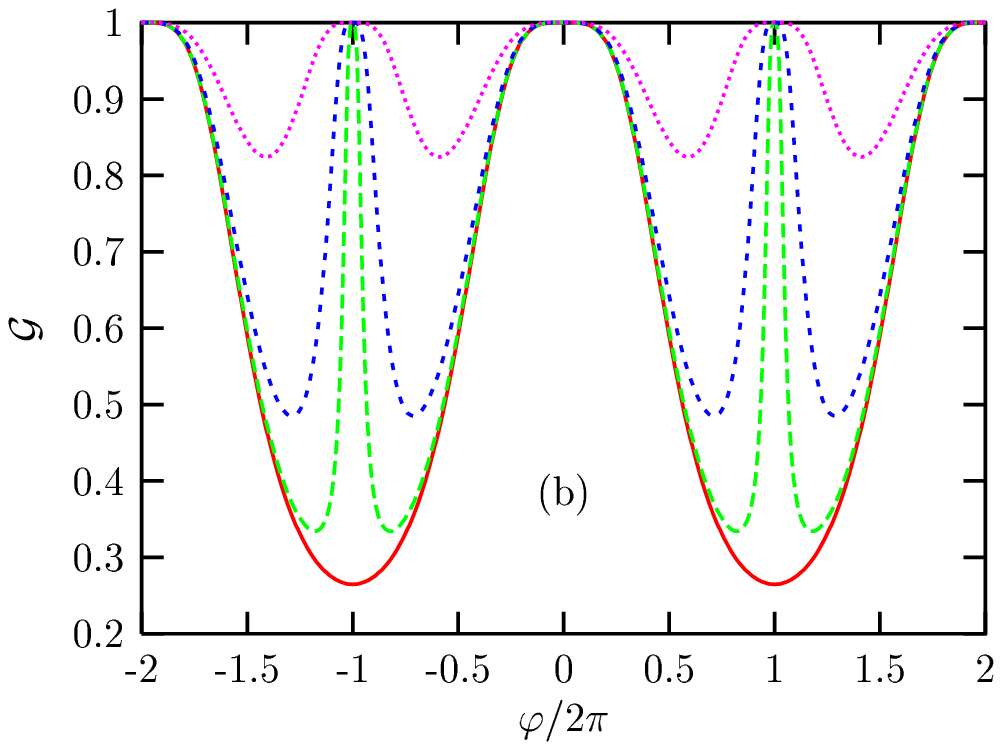}
\epsfxsize=1.6in
\epsffile{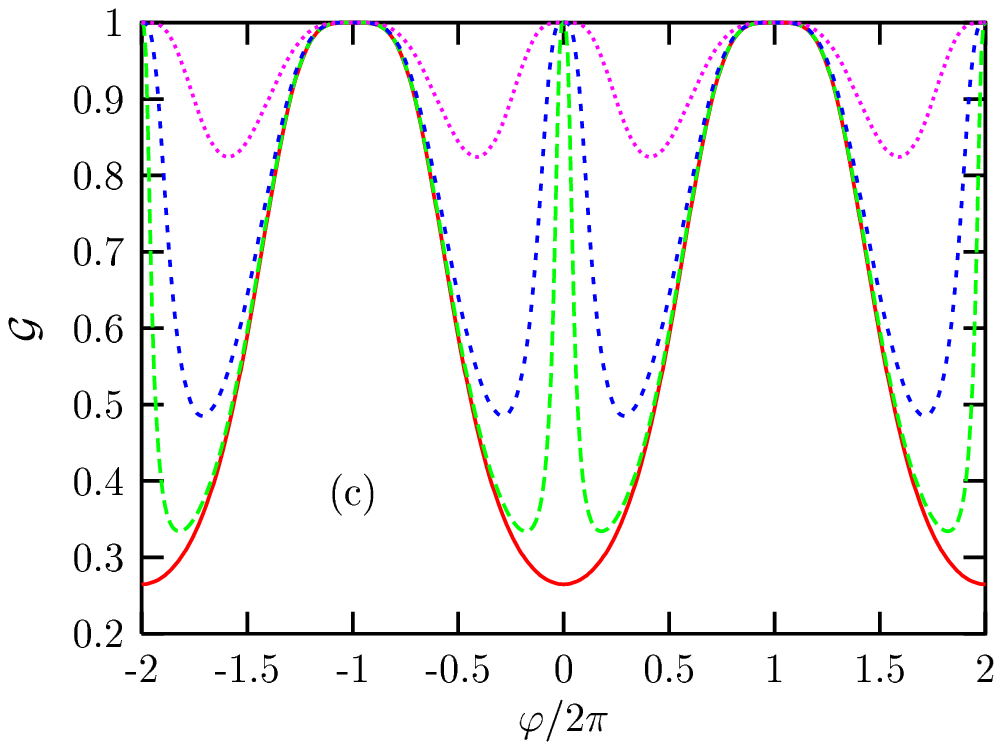}
 \caption{(a) Molecular two level energies as a function of the difference
  between the energy levels of the quantum dots. (b) AB oscillations for the
  bonding states marked in (a) by $S0$ (solid line), $S1$ (long-dashed line), 
  $S2$ (short-dashed line), $S3$ (dotted line). (c)  AB oscillations for the
  antibonding states marked in (a) by $A0$ (solid line), 
  $A1$ (long-dashed line), $A2$ (short-dashed line), $A3$ (dotted line).
    }
  \label{fig4}
\end{figure}
%
\end{document}